# Do CVD grown graphene films have antibacterial activity on metallic substrates?

L. Dellieu[1,*,§], E. Lawarée[2,§], N. Reckinger[1,3§], C. Didembourg[2], J.-J. Letesson[2], M. Sarrazin[1,3], O. Deparis[1], J.-Y. Matroule [2, $], and J.-F. Colomer[1,3,$]

[1]Solid-State Physics laboratory, Department of Physics, University of Namur, Rue de Bruxelles 61, 5000 Namur, Belgium.

[2]Research Unit in Microorganisms Biology, University of Namur, Rue de Bruxelles 61, 5000 Namur, Belgium.

[3]Research group on carbon nanostructures (CARBONNAGe), University of Namur, Rue de Bruxelles 61, 5000 Namur, Belgium.

[§, $] These authors contributed equally to this work.

Accurate assessment of the antibacterial activity of graphene requires consideration of both the graphene fabrication method and, for supported films, the properties of the substrate. Large-area graphene films produced by chemical vapor deposition were grown directly on copper substrates or transferred on a gold substrate and their effect on the viability and proliferation of the Gram-positive bacteria *Staphylococcus aureus* and the Gram-negative bacteria *Escherichia coli* were assessed. The viability and the proliferation of both bacterial species were not affected when they were grown on a graphene film entirely covering the gold substrate, indicating that conductivity plays no role on bacterial viability and graphene has no antibacterial activity against *S. aureus* and *E. coli*. On the other hand, antibacterial activity was observed when graphene coated the copper substrates, resulting from the release of bactericidal cupric ions in inverse proportion to the graphene surface coverage.

[*]Corresponding author : Email address : louis.dellieu@unamur.be (L. Dellieu)
FAX : +32 81 72 44 64

# 1. Introduction

Graphene is a two-dimensional crystal composed of hybridized-sp² carbon atoms in a hexagonal lattice structure that possesses many exceptional properties. *Inter alia*, it is a zero-gap semiconductor, mechanically hard, extremely flexible, chemically inert, impermeable to any atom or molecule, and optically transparent [1]. Such properties allow the development of biomedical devices with the proviso that the observed or suspected interactions between biological systems and graphene are well understood. In this context, a few toxicity studies have recently been conducted on graphene, in particular regarding its antibacterial activity, which have led to conflicting results in some cases [2-8]. One of the main reasons of the controversy is the nature and the properties of the various graphene-based materials. Graphite, graphene oxide, reduced graphene oxide or chemical vapor deposition (CVD) produced graphene can lead to dramatically different results when their impact on biological systems is studied. Most of the studies have been carried out on graphene oxide [9-11], which requires a subsequent reduction step for converting it into graphene [12]. The proliferation of L-929 cells [2] or neuroendocrine PC12 cells [3] on graphene paper resulting from the reduction of graphene oxide suggests the biocompatibility of this material. On the contrary, other results have shown that, on such a paper, the membrane integrity of *Escherichia coli* (*E. coli*) DH5 strain was lost, which induced a bactericidal effect [4]. Graphene oxide nanowalls, because of their very sharp edges, damage cell envelopes during the contact interaction with both Gram-negative bacteria *E. coli* and Gram-positive bacteria *Staphylococcus aureus* (*S. aureus*) and exhibit an even stronger antibacterial activity when they are in a reduced form [5]. Recently, this phenomenon was explained by the penetration of graphene oxide nanowalls in the membranes of *E. coli* resulting in the extraction of phospholipids and thereby in a reduction of the bacterial viability [6]. The antibacterial activity towards *E. coli* was also shown to depend on the type of graphene materials, i.e. increased activity from graphite to reduced graphene oxide [7]. Conversely, Ruiz *et al.* demonstrated that graphene oxide supports *E. coli* growth and therefore does not possess antibacterial properties [8]. These conflicting conclusions most likely come from the structural and chemical characteristics of the studied graphene materials, mainly graphene oxide nanowalls, which have sharp edges and must undergo a reduction step involving the presence of chemical functions on their surfaces. The antibacterial activity of structurally flat graphene films, on other hand, is an interesting avenue to investigate.

Very recently, large-area graphene films produced by CVD on Cu or Ge, or transferred to $SiO_2$ were reported to possess an antibacterial activity that was thought to be related to the electronic properties of the substrate [13]. The antibacterial mechanism was hypothesized to involve electron transfer from the microbial membrane, via graphene, to the substrate which acts as an electron pump [13]. In the present study, we provide experimental evidence that rules out the electron transfer model as a mechanism explaining the presumed antibacterial activity of CVD graphene films. This calls for revisiting the problem in order to clarify the role of the substrate conductivity, keeping in mind that any study on antibacterial activity must meet the required standards in biology in order to be convincing [14]. For this purpose, we elaborated CVD graphene films on copper (Cu) and gold (Au) substrates. Either entirely or partially covering films were grown on Cu surfaces whereas entirely covering films were transferred on Au surfaces. The antibacterial activity of the different samples was tested on the Gram-positive bacteria *S. aureus* and the Gram-negative bacteria *E. coli*.

## 2. Experimental details

### *2.1 Graphene samples*

The graphene samples were grown by atmospheric pressure CVD on Cu foil pieces [15,16] (~1 cm$^2$; 99.9% purity; 50-µm thick) with dilute $CH_4$ (95:5 Ar:$CH_4$) as carbon precursor, using a hotwall furnace and a quartz reactor. The Cu foils were cleaned by sonication in acetone and isopropanol, and Cu was de-oxidized with acetic acid. The synthesis was next conducted at a temperature of 1050 °C for 1 h under flows of Ar (500 sccm), $H_2$ (20 sccm), and dilute $CH_4$ (0.2 sccm for partial coverage and 1 sccm for full coverage). After graphene growth, the quartz tube was extracted rapidly under flows of Ar and $H_2$ in order to maintain the integrity of graphene [16]. Just after the synthesis, some Cu foils were covered by poly(methyl methacrylate) (PMMA) and etched by aqueous ammonium persulfate for subsequent transfer on Au substrates. After a few hours in the solution, the Cu foils were completely etched. The PMMA/graphene stacks were then copiously rinsed in distilled water in order to remove contaminants arising from the Cu etching step. Next, they were scooped from the solution with Au substrates and left to dry in air. After a second coating of PMMA [17], the samples were soaked in acetone for several hours to dissolve PMMA, rinsed in isopropyl alcohol and finally gently blown dry with nitrogen.

In order to verify the quality and the number of layers of the graphene films, they were transferred on Si/$SiO_2$ wafers and analyzed by Raman spectroscopy (see Fig. S1 in the supplementary information). Raman spectroscopy was performed at room temperature with a LabRam Horiba spectrometer at a laser wavelength of 514 nm. Figure S1a shows optical microscope image of a graphene layer after transfer on a Si/$SiO_2$ piece. The main background layer is typically monolayer graphene (Raman spectrum in Fig. S1b: 2D peak position = 2693 cm$^{-1}$; G peak position = 1590 cm$^{-1}$; $I_{2D}/I_G$ = 2.7; full width at half maximum = 33 cm$^{-1}$), with some bilayer islands (Raman spectrum in Fig. S1c: 2D peak position = 2713 cm$^{-1}$; G peak position = 1592 cm$^{-1}$; $I_{2D}/I_G$ = 0.9). In addition, the 2D peak of the bilayer graphene (full width at half maximum = 53 cm$^{-1}$) can be fitted by four lorentzians (see Fig. S1c), thereby revealing that the analyzed bilayer graphene islands are AB-stacked. All these values are in good agreement with the literature [18,19] and our previous work [16]. The defect-related band (located at ~1350 cm$^{-1}$) is barely visible, testifying to the good quality of the transferred graphene films.

It is noteworthy that graphene oxide may lead to bactericidal effects [10]. In order to ensure that this effect can be excluded in our case, we have assessed the possibility of reactive oxygen species generation by the graphene sheets. We inspected graphene on Cu foils by X-ray photoelectron spectroscopy (see reference [16] for more details) and found that the corresponding graphene film was not oxidized (within the detection limit of the equipment), as illustrated by C 1*s* spectrum in the supplementary information Fig. S2 (in accordance with our previous results [16]). Therefore it is very unlikely that our graphene sheets generate reactive oxygen species.

## *2.2 Bacterial live/dead analysis by flow cytometry*

After sterilization in a 75% v/v ethanol solution, triplicates of each surface (1 cm$^2$) were placed in a 24-well plate (NuncTM). In order to reduce water evaporation from the bacterial solution, the remaining wells and the empty spaces between them were filled with deionized water. Sixty µl of a bacterial suspension containing 6×10$^6$ CFU/ml of *S. aureus* (ATCC 25923) or 2×10$^7$ CFU/ml of *E. coli* (ATCC 25922) were poured onto the surfaces and incubated for 24 h at 37 °C. In order to assess bacterial viability, bacteria were diluted in a 0.85% NaCl solution and stained with the LIVE/DEAD® BacLight$^{TM}$ Bacterial Viability Kit (Molecular Probes®) according to the manufacturer protocol. 50,000 events per experimental condition were acquired with a BD FACSCalibur™ flow cytometer equipped with an argon laser at 488 nm. Data acquisition and analysis were performed by using the BD CellquestPro™ software. The bacterial population was distributed in two regions of the log-integrated red (Propidium iodide:FL3) versus green (Syto9:FL1) plot and the number of bacteria found in these regions were used to estimate the percentage of bacterial viability. The non-bacterial events were eliminated by gating the SSC/FSC (granularity against cell size) plot. To ensure the statistical validity of the results, each experiment was repeated three times (biological replicate) and each biological replicate was analyzed three times (technical replicate).

## 2.3 Measurement of Cu and Au concentration by atomic absorption spectroscopy

The metallic ions concentration in the bacterial suspensions was measured by using atomic absorption spectroscopy (AAS) (AA-7000F from Shimadzu). The AAS was calibrated with different standard solutions of the analysed element. Bacterial suspensions were diluted 100 times in 1 M $HNO_3$. The Cu or Au absorbance of each biological replicate was measured three times as technical replicates.

## 2.4 Evaluation of the antibacterial activity

The bacterial suspension, incubated for 24 h on the surfaces described above, was collected and serially diluted in liquid LB medium. Twenty µl of each dilution were inoculated on LB agar plates. After an overnight incubation at 37 °C, the culture plates were photographed and the total bacterial charge was estimated by counting the Colony Forming Units (CFU). To ensure the statistical validity of the results, the experiment was repeated three times (biological triplicate). Each biological replicate was inoculated on three different LB agar plates (technical replicate).

## 2.5 Scanning electron microscopy analysis

Scanning electron microscopy (SEM) was performed using a JEOL 7500 F microscope under an accelerating voltage of 1 kV, with a working distance of 3 mm, and an emission current of 5 µA. The samples were not coated with any metallic thin layer.

## 3. Results and discussion

### 3.1 Production and characterization of the graphene/metal samples

The use of a non-biologically inert metallic substrate, such as Cu, which was previously used to test graphene antibacterial activity [13] is expected to bias the assessment of this activity considering the bactericidal and bacteriostatic properties of Cu [20]. As opposed to Cu, Au is barely toxic for bacterial or animal cells due to its elemental properties [21]. Therefore, Au is an excellent candidate to test the antibacterial activity of graphene films on metallic substrates and to assess the contribution of the substrate conductivity. Both substrates have been considered in the present study.

As already mentioned here above, the graphene films were synthesized by atmospheric pressure CVD on Cu foils [15,16] to produce two types of samples. The first one, named graphene@Cu, entirely covers the Cu surface and is composed of micrometer-sized domains of predominantly monolayer graphene with some bi- and trilayer graphene islands. SEM images (Fig. 1a-b) show a graphene film entirely covering the Cu substrate. The low magnification SEM image (Fig. 1a) is poorly contrasted because graphene completely overlays the Cu surface. In Fig. 1b, areas with different contrasts are seen: the darker the contrast, the greater the number of layers. Wrinkles, folds in graphene due to the difference between the thermal expansion coefficients of Cu and graphene, are also seen to cross Cu grain boundaries (Insert to Fig. 1b), testimony to the continuity of the graphene film [22].

The second type of sample, named graphene$^{+/-}$@Cu, is composed of single hexagonal domains that merge to form larger ones. In stark contrast with Fig. 1a-b, the low magnification SEM image (Fig. 1c) reveals a strong contrast between Cu and graphene because graphene hexagonal domains, about one to two tens of microns in size, do not fully cover the Cu foil. Moreover, these domains are not only composed of graphene monolayer but also occasionally of bi- or trilayer flakes (Fig. 1d). Graphene surface coverage was estimated to be around 80% from SEM image analysis.

Another type of sample, named graphene@Au, consists of continuous films of graphene deposited on Au substrates by transfer. Graphene was obtained exactly in the same CVD synthesis conditions as for graphene@Cu and was transferred on Au substrates by using the polymer-assisted technique [23]. Finally, bare Cu and Au substrates are used as control samples.

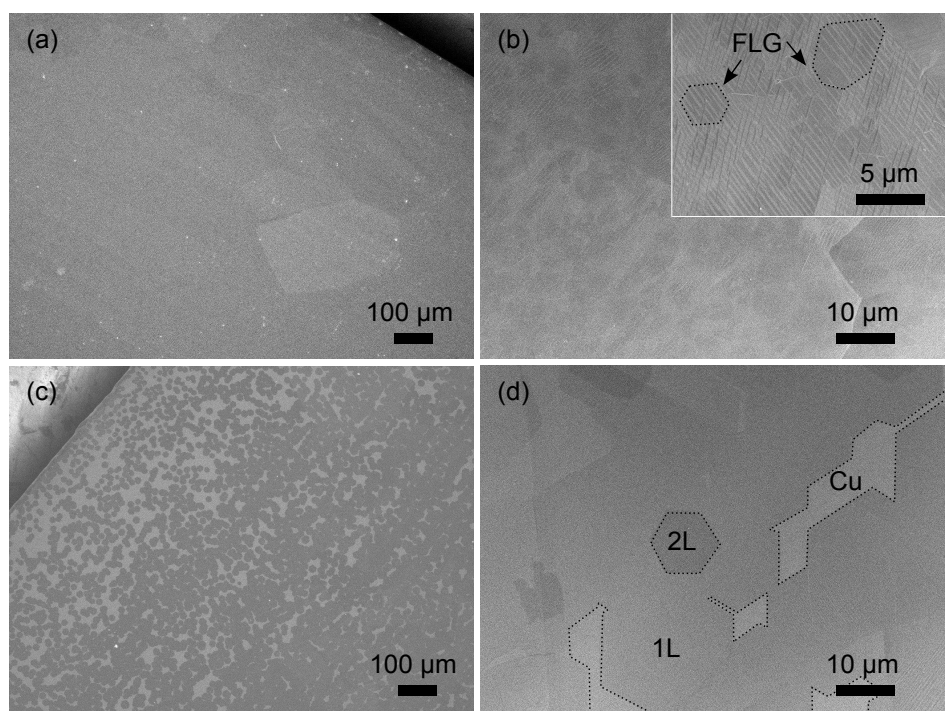

**Fig 1: Morphology of graphene on Cu**. (a) Low magnification SEM image of graphene@Cu showing the complete coverage of graphene on the Cu surface. (b) Medium magnification SEM image of graphene@Cu. Insert: Higher magnification SEM image where few-layer graphene domains are highlighted. The contrast differences are due to graphene domains with different numbers of layers, Cu grain orientations and boundaries, or graphene wrinkles. (c) Low magnification SEM image of graphene$^{+/-}$@Cu showing graphene hexagonal domains (dark grey islands) on the Cu substrate (light grey background), evidencing the partial coverage of the Cu surface. (d) Medium magnification SEM image of graphene$^{+/-}$@Cu showing the contrast difference between the graphene domains and the Cu substrate, and also between different numbers of graphene layers (1L: monolayer, 2L: bilayer).

## *3.2 Graphene has no antibacterial activity.*

The bacterial envelope is a finely regulated physical barrier between the intracellular and extracellular environments. Severe structural alterations of this barrier are usually non-reversible and lead to bacterial cell death. Graphene bactericidal properties were assessed in the Gram-positive *S. aureus* and the Gram-negative *E. coli* with the so called "LIVE/DEAD" assay which relies on the level of bacterial envelope integrity to discriminate live from dead bacteria. Combination of the LIVE/DEAD assay with flow cytometry provides a quantitative insight of bacterial cell death (see Fig. S3 in supplementary information). *S. aureus* or *E. coli* cultures were seeded into a 24-well plate containing (i) bare Cu or bare Au, (ii) a Cu foil partially (graphene$^{+/-}$@Cu) or fully covered (graphene@Cu) with graphene or (iii) an Au foil fully covered with graphene (graphene@Au). After a 24 h incubation at 37 °C, the cell cultures were recovered and subjected to the LIVE/DEAD assay. Interestingly, *S. aureus* and *E. coli* grown on graphene@Cu or graphene@Au exhibit at least 91% viability relative to the 97% and 94% viability of the control *S. aureus* and *E. coli* cultures grown directly on the bottom of the 24-well plate (Fig. 2 a-b). Statistical analysis reveals that the difference between graphene@Cu condition (but not the graphene@Au condition) and the negative control is significant (p-value of 0.0004 for *S. aureus*, p-value of 0.0106 for *E. coli*). Accordingly, only 66% of *S. aureus* and 54% of *E. coli* could survive when grown on graphene$^{+/-}$@Cu. *S. aureus* cells grown on bare Cu all died whereas 6% of *E. coli* cells survived under the same conditions (Fig. 2 a-b). AAS performed on the *S. aureus* and *E. coli* culture medium after the 24 h incubation revealed a ~19 mM, ~6 mM and ~0.2 mM Cu concentration for bare Cu, graphene$^{+/-}$@Cu and graphene@Cu, respectively (Fig 2 c-d), demonstrating the release of cupric ions into the culture medium and, therefore, the relative permeability of the graphene sheets. Indeed, graphene was previously shown to be an imperfect barrier owing to the presence of defects and domain boundaries [24,25]. Au concentration in the culture medium remained very low (~0.8 µM) under any experimental conditions, likely due to its low solubility.

Together, these results suggest that graphene *per se* has no bactericidal activity. Nevertheless, one cannot rule out any bacteriostatic effect of graphene films, where *S. aureus* and *E. coli* cells would survive to graphene in a non-replicative form. In order to test this hypothesis, we performed a CFU assay which quantitatively measures the ability of bacteria to divide and to give macroscopic colonies (see Fig. S4 in supplementary information) [26]. Twenty-four hours *S. aureus* and *E. coli* cultures performed under the experimental conditions described above were serially diluted and seeded on agar culture plates which were further incubated overnight until CFUs counting.

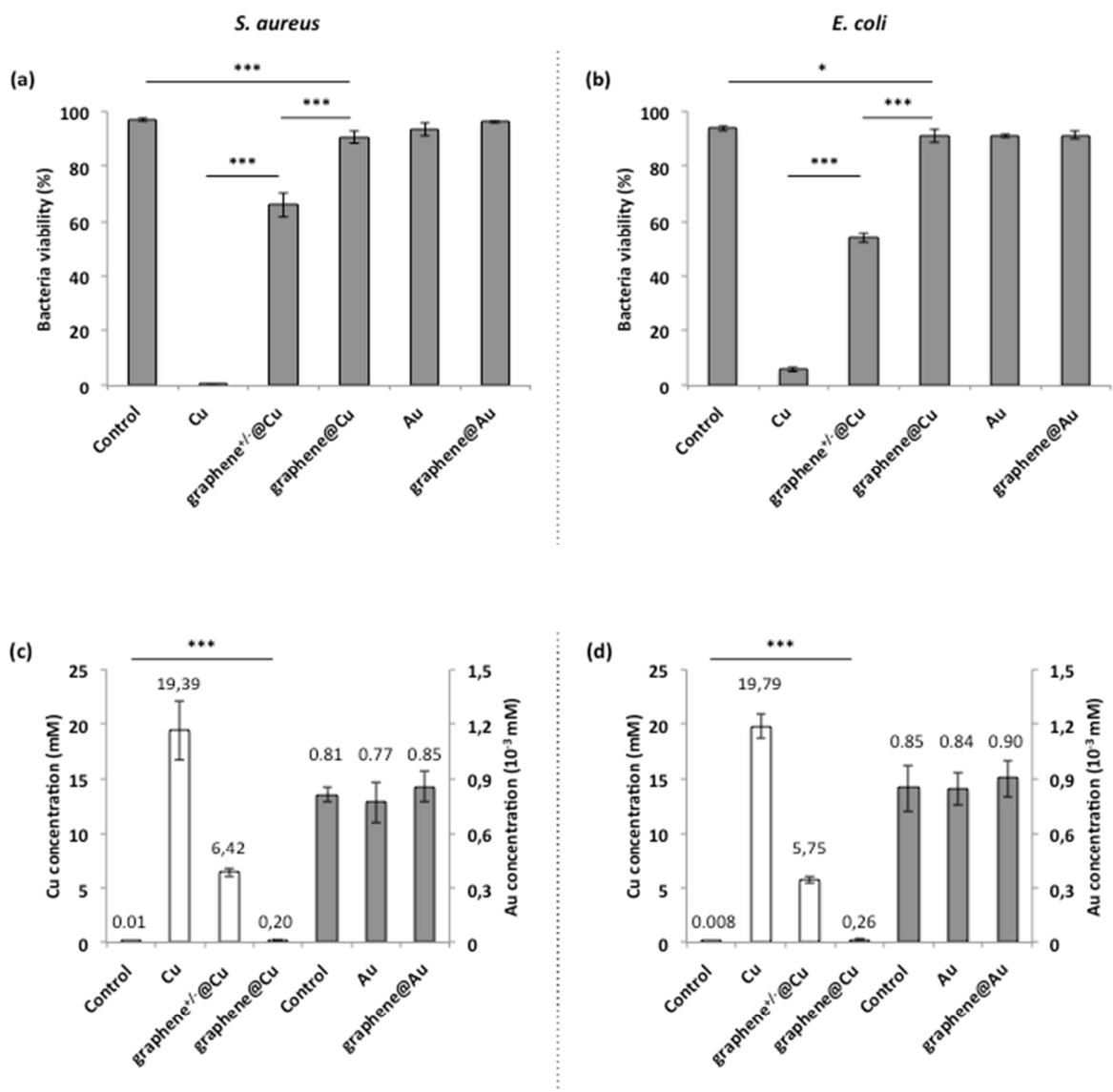

**Figure 2: Graphene *per se* has no impact on *S. aureus* and *E. coli* viability.** Bacterial viability was determined using the LIVE/DEAD assay after a 24 h incubation of *S. aureus* (a) and *E. coli* (b) on bare Cu, bare Au, graphene$^{+/-}$@Cu, graphene@Cu and graphene@Au. Cu and Au concentrations were measured by AAS after incubating *S. aureus* (c) and *E. coli* (d) for 24 h under the same experimental conditions. Stars denote the statistical significance (p-values) in increasing order: p<0.1 (*), p<0.05(**), p<0.01(***). Error bars indicate standard deviation values.

During the 24 h incubation in the 24-well plate, the control *S. aureus* and *E. coli* cultures underwent a 70000 and 700 fold increase of the CFU/ml, respectively, indicating that both bacterial species are able to proliferate under these conditions (Fig. 3 a-b). A similar pattern of proliferation was observed when both *S. aureus* and *E. coli* were grown on graphene@Au corroborating the lack of cytotoxicity of the graphene@Au evidenced by the LIVE/DEAD assay. When grown on graphene$^{+/-}$@Cu or graphene@Cu, both *S. aureus* and *E. coli* cells displayed a decrease of their proliferation rate, likely due to the toxicity of the released cupric ions. Logically, no CFU was observed when *S. aureus* was grown on bare Cu, which did not completely abolish the capacity of *E. coli* to proliferate though since 0.01% of the seeded *E. coli* cells could give rise to a colony. These 0.01% are still much below the 6% survival measured with the LIVE/DEAD assay (Fig. 2 b), suggesting a bacteriostatic effect of the released cupric ions on *E. coli*. We therefore conclude that graphene has no significant intrinsic antibacterial activity when it is deposited on a biologically inert metallic substrate.

*3.4 Discussion*

Our results demonstrate that graphene, as far as it is produced by CVD and deposited on a conductive (metallic) substrate, does not display any significant antibacterial property against *S. aureus* and *E. coli*. Our study refutes the electron transfer model recently proposed by Li *et al.* to explain the antibacterial activity of large-area graphene films coated on Cu [13]. From a physical point of view, Li *et al.*'s model is questionable because it neglects the fact that graphene/metal junction properties must be carefully considered when estimating work functions of materials [27-34]. First, when graphene rests on Cu, it is known that it should be usually *n*-doped [27-31]. That means that electrons tend to be transferred from Cu to graphene, and not the opposite way. Consequently, graphene@Cu should hardly allow electron transfer from graphene to Cu unlike what is assumed in Li *et al.*'s model. In fact, considering the theoretical current-voltage characteristic of graphene@Cu, the result predicted by Yamacli [29] is just the opposite of the one expected by Li *et al.*'s assumption. By contrast, graphene@Au is *p*-doped [29-34] and therefore electrons tend to be transferred from graphene to Au. In this case, the theoretical current-voltage characteristic of graphene@Au [29] meets the assumption of Li *et al.*'s model in terms of electron transfer direction. Now, as a matter of fact, no antibacterial activity was observed experimentally with graphene@Au although electron transfer from graphene to Au is well known to occur efficiently [29-34]. Therefore, we can conclude that the proposed electron transfer mechanism is not pertinent as far as the interaction between bacteria and graphene on conductive substrates is concerned.

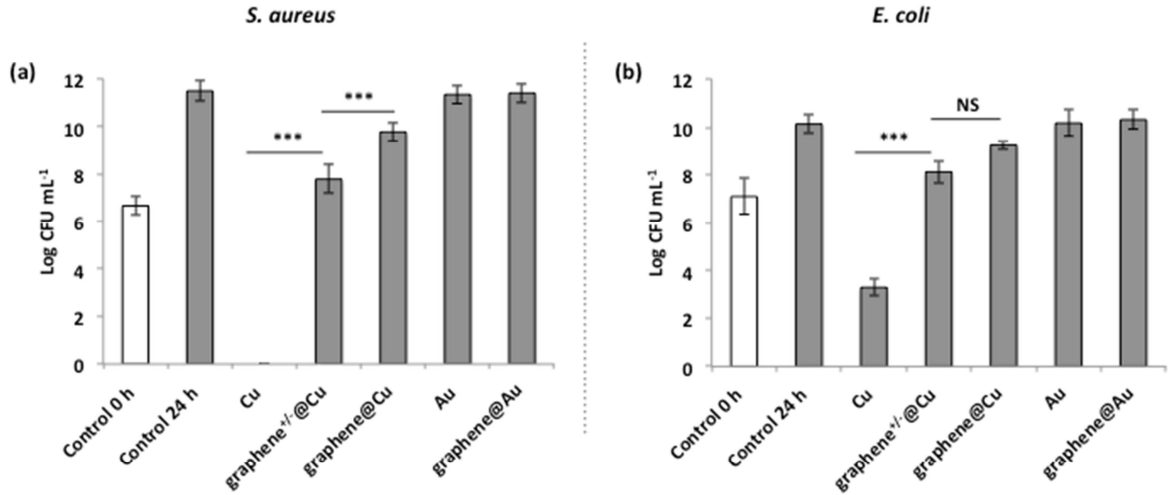

**Fig. 3: Graphene per se has no antibacterial activity on *S. aureus* and *E. coli*.** The ability of *S. aureus* (a) and *E. coli* (b) to proliferate upon a 24 h incubation (black bars) on the plastic bottom (control 24 h), on bare Cu, bare Au, graphene$^{+/-}$@Cu, graphene@Cu and graphene@Au was assessed by counting the CFU obtained after serially diluting and overnight plating the bacteria on solid medium. One control bacteria sample (control 0 h) was also immediately subjected to CFU counting with no 24 h incubation (white bar) in order to determine the proliferation of the control bacteria. Stars denote the statistical significance (p-values) in increasing order: p<0.1 (*), p<0.05(**), p<0.01(***). Error bars indicate standard deviation values. NS : non significant.

The electron transfer model is also highly questionable from a biological point of view because, according to the Li *et al.*, it involves an electron transfer from the bacterial respiratory chain to an external electron acceptor through an "electron conduit". Respiration in bacteria is an energy generating process that relies on electron transport using a chain of proteins located in the cytoplasmic membrane to a terminal electron acceptor. This transport, coupled to the extrusion of protons ($H^+$), leads to a charge and pH gradient through the membrane. These charge carriers are far from being in contact with the external medium. In most cases, this does neither involves an extra-cellular electron acceptor nor a bacterial nanowire as claimed in Li *et al.*'s article. With regard to *S. aureus*, in aerobic conditions, the respiratory chain uses molecular $O_2$ as a final acceptor and this gas freely diffuse to the cytoplasmic membrane [35]. According to our results, the hypothesis of facile transfer of electrons from microbial membranes to graphene does not seem satisfactory at all. Indeed, we do not observe any antibacterial effect with graphene@Au surfaces, despite that the Au-graphene junction should favor such a hypothetical transfer. Notwithstanding that the Cu-graphene junction should not facilitate electron transfer, we observe here an antibacterial activity due to released cupric ions which is in inverse proportion to graphene surface coverage. Indeed, the bactericidal property of cupric ions must be taken into account since the dissolved cupric ions cause membrane damage and cell disintegration [37].

# 4. Conclusion.

The absence of antibacterial activity of large-area CVD graphene films on conductive (Au, Cu) substrates was demonstrated for *S. aureus* and *E. coli*. This result implies that the conductive character of the substrate has no influence on the viability of *S. aureus* and *E. coli* bacteria in contact with CVD graphene films. When a Cu substrate is used, however, the release of cupric ions from areas not covered by the graphene film led to an antibacterial effect which depends on the degree of graphene coverage. This role of the Cu substrate should be carefully taken into account in any assessment study on graphene antibacterial activity.

**Acknowledgments.**

A part of this work is financially supported by the Belgian Fund for Scientific Research (F.R.S.-FNRS) under FRFC contract ''Chemographene'' (No. 2.4577.11). L. D. and E. L. are supported by the Belgian Fund for Industrial and Agricultural Research (FRIA). J.-F. Colomer is supported by the F.R.S.-FNRS Research Associate. This research used resources of the Electron Microscopy Service located at the University of Namur ("Plateforme Technologique Morphologie – Imagerie"). The authors thank Carole Morel, Pierre Cambier and Eloise Van Hooijdonk for their technical support and advice. The authors acknowledge Xiaohui Tang and Benoît Hackens for their help with Raman measurements.

# Supplementary information

# Do CVD grown graphene films have antibacterial activity on metallic substrates?

L. Dellieu[2,*,§], E. Lawarée[2,§], N. Reckinger[1,3§], C. Didembourg[2], J.-J. Letesson[2], M. Sarrazin[1,3], O. Deparis[1], J.-Y. Matroule[2,$], and J.-F. Colomer[1,3,$]

[1]Solid-State Physics laboratory, Department of Physics, University of Namur, Rue de Bruxelles 61, 5000 Namur, Belgium.

[2]Research Unit in Microorganisms Biology, University of Namur, Rue de Bruxelles 61, 5000 Namur, Belgium.

[3]Research group on carbon nanostructures (CARBONNAGe), University of Namur, Rue de Bruxelles 61, 5000 Namur, Belgium.

[§, $] These authors contributed equally to this work.

[*]Corresponding author : Email address : louis.dellieu@unamur.be (L. Dellieu)
FAX : +32 81 72 44 64

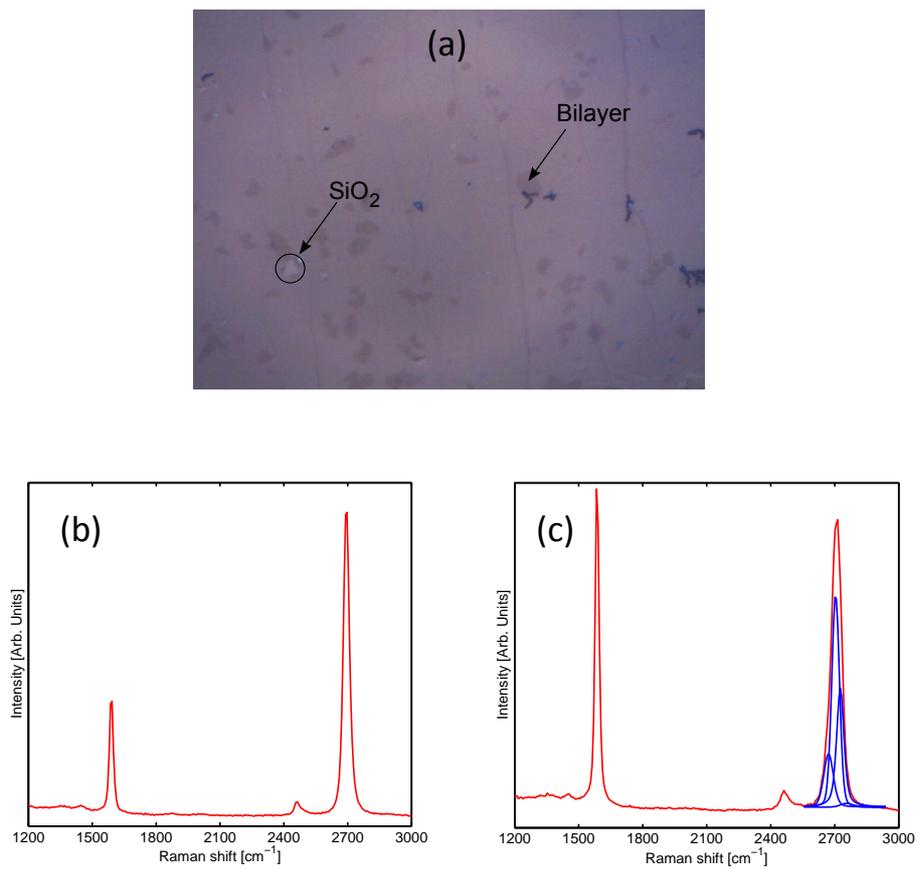

**Fig. S1** – (a) Optical image of a graphene film transferred onto a Si/SiO$_2$ substrate. Corresponding Raman spectra for (b) monolayer and (c) bilayer graphene, respectively. The 2D peak of bilayer graphene can be fitted with 4 lorentzians, indicating that it is AB-stacked.

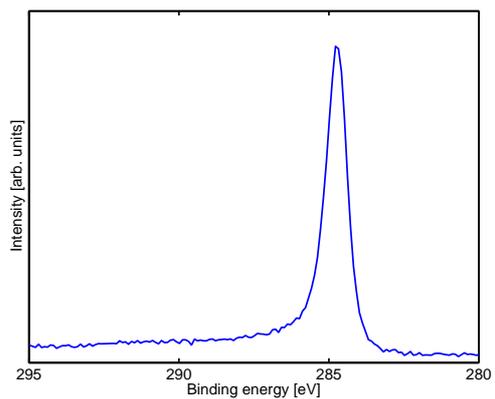

**Fig. S2** – C 1*s* X-ray photoelectron spectroscopy spectrum of graphene grown on copper foils, showing that graphene is not oxidized.

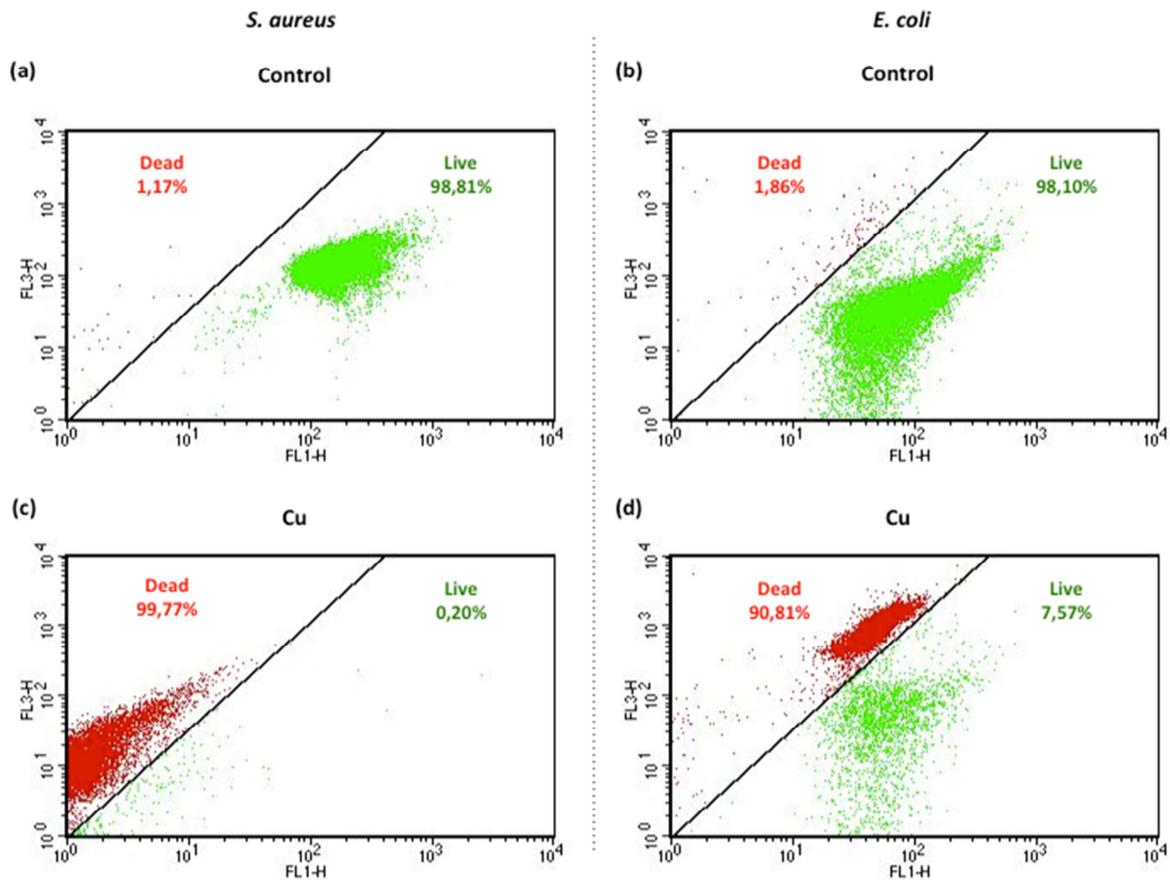

**Fig. S3.** Flow cytometric dotplot analysis of bacteria stained with the LIVE/DEAD assay. *S. aureus* (a, c) and *E. coli* (b, d) were incubated for 24 h on the plastic bottom (a, b) or on bare Cu (c, d). Each dot of the plot represents a single bacterium which is stained according to its physiological status: red for severe membrane damage (death), green for no detectable membrane damage (presumably alive). Y-axis: red fluorescence with propidium iodide, X-axis: green fluorescence with syto 9.

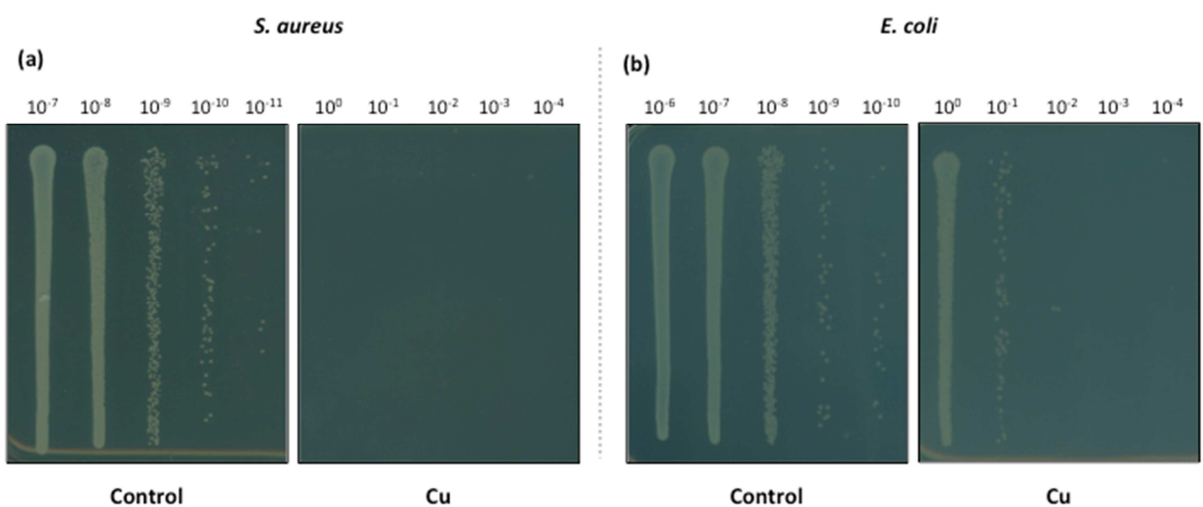

**Fig. S4**. Typical photographs of serially diluted *S. aureus* (a) and *E. coli* (b) bacteria grown on agar culture plates after a 24 h incubation on the plastic bottom (control) or on bare Cu.